\documentclass{mn2e}
\usepackage{times,epsfig}
\newcommand{\aap}{A\&A}
\newcommand{\aaps}{A\&AS}
\newcommand{\aj}{AJ}
\newcommand{\apj}{ApJ}

\newcommand{\apjs}{ApJS}

\newcommand{\mnras}{MNRAS}
\newcommand{\pasp}{PASP}

\def \kms{\ifmmode{~{\rm km\,s}^{-1}}\else{~km~s$^{-1}$}\fi}
\def \vhel{\ifmmode{V_{{\rm hel}}}\else{$V_{{\rm hel}}$}\fi}
\def \vsys{\ifmmode{V_{{\rm sys}}}\else{$V_{{\rm sys}}$}\fi}
\def \vobs{\ifmmode{V_{{\rm obs}}}\else{$V_{{\rm obs}}$}\fi}
\def \degree{\ifmmode{^{\circ}}\else{$^{\circ}$}\fi}
\def \lsun{\ifmmode{{\rm\ L}_\odot}\else{${\rm\ L}_\odot $}\fi}
\def \msun{\ifmmode{{\rm\ M}_\odot}\else{${\rm\ M}_\odot$}\fi}
\def \myr{\ifmmode{{\rm\ M}_\odot{\rm\ yr}^{-1}}\else{${\rm\ M}_\odot$ 
yr$^{-1}$}\fi}
\def \teff{\ifmmode{{\rm{T}}_{\rm eff}}\else{${\rm{T}}_{\rm eff}$}\fi}
\def \mdot{\ifmmode{{\rm\dot{M}}}\else{${\rm\dot{M}}$}\fi}

\newcommand{\ha}{H$\alpha$}

\newcommand{\heii}{He\,{\sc ii}}
\newcommand{\heiil}{He\,{\sc ii}\ 6560\,\AA}
\newcommand{\heiis}{He\,{\sc ii}\ 4686\,\AA}

\newcommand{\oiii}{[O\,{\sc iii}]}
\newcommand{\oiiil}{[O\,{\sc iii}]\ 5007\,\AA}
\newcommand{\nii}{[N\,{\sc ii}]}
\newcommand{\niil}{[N\,{\sc ii}]\ 6584\,\AA}

\newcommand{\niiab}{[N\,{\sc ii}]\ 6548,\ 6584\,\AA}
\def \st{\ifmmode{^{\mathrm{st}}}\else{${^{\mathrm{st}}}$}\fi}
\def \nd{\ifmmode{^{\mathrm{nd}}}\else{${^{\mathrm{nd}}}$}\fi}
\def \rd{\ifmmode{^{\mathrm{rd}}}\else{${^{\mathrm{rd}}}$}\fi}
\def \th{\ifmmode{^{\mathrm{th}}}\else{${^{\mathrm{th}}}$}\fi}
\newcommand{\hnii}{{\rm H}$\alpha+$[N {\sc ii}]}

\title[Helix planetary nebula]{The creation of 
the Helix planetary nebula (NGC~7293) by
multiple events}
\author[J. Meaburn et al]
{J. Meaburn$^{1}$, P. Boumis$^{2}$, 
J. A. L\'{o}pez $^{1}$,
 D. J. Harman$^{3,}$ $^{4}$,
 M. Bryce $^{3}$,\newauthor
M.P. Redman$^{5}$  and F. Mavromatakis $^{6}$.\\ 
$^{1}$ 
Instituto de Astronom\'{\i}a, UNAM, Apdo. Postal 877,
Ensenada, B.C. 22800, M\'{e}xico.\\
$^{2}$Institute of Astronomy \& Astrophysics, National Observatory of
Athens, I. Metaxa \& V. Paulou, GR--152 36 P. Penteli, Athens,
Greece\\
$^{3}$Jodrell Bank Observatory, Dept of Physics \& Astronomy, University of
Manchester, Macclesfield, Cheshire SK11 9DL UK.\\
$^{4}$Liverpool John Moores University, Birkenhead, CH41 1LD, UK\\
$^{5}$ Department of Physics, National University of 
Ireland Galway, Galway, Ireland. \\
$^{6}$ University of Crete, Physics Department, P.O. Box 2208, 710 03
Heraklion, Crete, Greece.
}
\begin{document}

\date{Accepted 2005 April 5. Received 2005 March 25; in original 
form 2005 January 6}

\pagerange{\pageref{firstpage}--\pageref{lastpage}} \pubyear{2005}

\maketitle

\label{firstpage}

\begin{abstract}

A deep, continuum--subtracted, image of NGC 7293 has
been obtained in the light of the \hnii\ emission lines. New images
of two filamentary halo stuctures have been obtained 
and the possible detection of a collimated
outflow made.

Spatially resolved, longslit profiles of the \hnii\ lines have
been observed across several of these features with the
Manchester  echelle spectrometer combined with the 
San Pedro Martir 2.1--m telescope; these are compared with 
the \niil, \oiiil, \heiil\ and \ha\ profiles obtained over the 
nebular core.

The central \heii\ emission is originating in a $\approx$~0.34 
pc diameter spherical
volume expanding at $\leq$ 12 \kms\ which is surrounded, and partially
coincident with an \oiiil\ emitting inner shell
expanding at 12 \kms. The bright helical structure surrounding this 
inner region is modelled as a bi--polar nebula with lobe expansions
of 25 \kms\ whose axis 
is tilted at 37\degr\  to the sight line but with a toroidal
waist itself expanding at 14 \kms.

These observations are compared with the expectations of the
interacting two winds model for the formation of PNe. Only
after the fast wind has switched off could this global
velocity structure be generated. Ablated flows must complicate any 
interpretation.

It is suggested that the clumpy nature of much of the material 
could play a part
in creating the radial `spokes' shown here to be apparently present
close to the central star. These `spokes' could
in fact be the persistant tails of cometary globules
whose heads have now photo--evaporated completely.

A halo arc projecting from the north--east of the bright core
has a conterpart to the south--east. 
Anomolies in the position--
velocity arrays of line profiles could suggest that these
are part of an expanding disc not aligned with the 
central helical structure though expanding bi--polar lobes
along a tilted axis are not ruled out.

\end{abstract}


\section{Introduction}

The multitude of phenomena in the  
Helix planetary nebula (NGC 7293, PK 36--57\degr1) at a distance of only
213 pc  (Harris et al 1997) can be investigated over 
a uniquely wide
range of spatial scales for at that distance 1\arcsec\ $\equiv$ 
3.19~$\times$~10$^{15}$ cm.
It is therefore proving to be one of the most 
important laboratories for the investigation
of all aspects of evolved planetary nebulae (PNe).

The progenitor star is a low luminosity
(L/\lsun\ $\approx$ 100 Henry, Kwitter \& Dufour 1999) 
white dwarf (WD 2226--210, Mendez et al 1988)
with an effective temperature of 117,000~K and mass 0.93~\msun\ 
(G\'{o}rny, Stasinska \& Tylenda 1997)
as well as a late--type 
dMe companion whose halo is responsible for hard X--ray emission 
(Chu et al 2004; Gruendl et al 2001). Patriarchi \& Perinotto (1991)
discovered that 60 percent of central stars of PNe emit particle winds
from 600 -- 3500 \kms\ but failed to detect one from WD 2226--210. In fact,
Cerruti--Sola \& Perinotto (1985) had given an upper limit to such a wind
as $\leq$~10$^{-10}$ \myr\ which has implications for the dynamics of NGC 7293
(see Sect. 4.1).
The lowly ionised, apparantly helical, structure that gives the
nebula its name could be a toroid with bi-polar lobes 
viewed at $\approx$~37\degr\ with respect to the bi--polar axis (
Meaburn et al
1998; Henry, Kwitter \& Dufour 1999) though O'Dell, McCullough \& Meixner
(2004) present an alternative view (Sect. 3.2). 
Whatever form this helical structure takes it contains an \oiii\
emitting  shell on the inside surface of the toroid with an inner
\oiii\ emitting shell (Meaburn \& White 1982; O'Dell et al 2004)
surrounding a highly ionised \heii\ emitting central spherical
volume shown particularly well by O'Dell et al (2004) and 
considered by Henry et al (1999). This 
bright central nebula is surrounded by faint filamentary structure
as shown in different ways by Malin (1982), Walsh \& Meaburn (1987),
Speck et al (2002) and O'Dell et al (2004).

The whole ionized nebula is enveloped within a massive envelope of molecular
gas (Storey 1984; Huggins \& Healey 1989; Healey \& Huggins 1990;
Forveille \& Huggins 1991; Young et al 1999).

Arguably the most interesting feature of the ionised helical structure
are the  multitude of `cometary' knots
inhabiting its inside surface. These have
dense neutral cores (Meaburn et al 1992, 1998; Huggins et al 1992)
with long ionised tails pointing radially away from the central star seen
in superb detail in the HST images of O'Dell \& Handron (1996).

In the present paper, spatially resolved profiles of the \ha\ and \niil\
emission lines are compared with new continuum--subtracted emission
line images of the most prominent halo feature. Furthermore,
the kinematics of the nebular core are explored with 
longslit profiles of the \oiiil\ and \heiil\ emission lines.The central motions
are then compared with previously published data
and the implications for current dynamical models for
the creation of PNe are considered.
Furthermore, the
bi--polar/toroidal model of the bright helical structure is tested
by numerical simulations compared in detail with observations.
The origin of strange radial `spokes' of enhanced \nii\ emission is also
explored. 

\section{Observations and Results}

\subsection{Wide -- field imagery}

The image of NGC 7293 shown in Figs. 1a--c was obtained
 with the 0.3--m (f/3.2) Schmidt-Cassegrain 
telescope at Skinakas Observatory, Crete, Greece in 2003 July 30. The 
1024$\times$ 1024
(19$\times$19 $\mu$m$^{2}$) pixels Thomson CCD camera was used
resulting in a scale of 4\arcsec.1 pix$^{-1}$~and a field
of view of 70\arcmin~$\times$~70\arcmin.
Five exposures in \hnii\ of 2400 s each were taken during
the observations (resulting in a total integration 
time of 12000 s), while three 
exposures of 240 s were obtained with the continuum filter. The latter were 
subtracted from the former to eliminate the confusing star field (more 
details of this technique can be found in Boumis et al 2002).
The image reduction 
(bias subtraction, flat-field correction) was carried out using the 
standard IRAF and MIDAS packages. The astrometry information was 
calculated using stars from the Hubble Space Telescope (HST) Guide
Star Catalogue (Lasker et al. 1999). Although many of the features
shown in Fig. 1 a) \& b) are previously known (see refs. in Sect. 1 
and particulary  O'Dell et al 2004) this unique wide--field 
continuum--subtracted image relates them to each other clearly.
The very deep presentation in Fig. 1 c) reveals a bow--shaped feature and 
possible, but tenuous, counter--jet both of which are arrowed.

\subsection{Longslit spectrometry}

\subsubsection{\ha\ \& \niil\ profiles from the halo}

\noindent 
The longslit observations were obtained with the Manchester Echelle
Spectrometer (MES - Meaburn et al, 1984 \& 2003)
combined with the f/7.9 focus of the 2.1--m San Pedro M\'{a}rtir UNAM
telescope on 28 June 1998.  This echelle spectrometer has no
cross-dispersion. For the present observations, a filter of 90~\AA\
bandwidth was used to isolate the 87$^{th}$ order containing the \ha\ and
\nii\ nebular emission lines.

A Tektronix CCD with 1024~$\times$~1024 square pixels,
each with 24~$\mu$m sides, was
the detector.  Two times binning was employed in both the spatial and
spectral dimensions. Consequently 512 increments, each 0.60\arcsec\
long, gave a total projected slit length of 5.12\arcmin\ on the sky.
`Seeing' varied between 1-2\arcsec\ during these observations.  The
slit was 150~$\mu$m wide ($\equiv$~11~\kms\ and 1.9\arcsec) and the
integration times were 1800 s. The spectra were calibrated in
wavelength to $\pm$~1~\kms\ accuracy against that of a Th/Ar arc lamp
and in absolute surface brightness (of the \ha\ line) to an accuracy
of $\pm$~20~percent against a slitless spectrum of the `standard' star
Feige 56.

The three slit positions (1--3) over the halo feature of NGC~7293
are shown in Fig. 2 against part of the image in Fig. 1. Greyscale
representations of the resultant position--velocity (pv) 
arrays of \ha\ and \nii\
profiles are shown in Figs.~3--5 respectively. 

\subsubsection{\heiil\ and \oiiil\ profiles from the core}

Previously (Meaburn et al 1996, 1998) \ha\ and \nii\ profiles
were obtained with MES on the Anglo--Australian (3.9--m) telescope (AAT), 
from NS and EW  lines of measurements, each 1000 arcsec long,
through the central star of NGC 7293. Spectra from many
slit lengths were combined to cover such large regions (see the
above papers for all technical details). For comparison with the
present spectral observations the pv array of the individual
velocity components in the \niil\ profiles along the EW line of measurements
is shown in Fig. 6 (from Meaburn et al 1996).

A re-examination of these previous pv arrays revealed that the  high 
excitation \heiil\ emission line, from an area
of $\approx$~200\arcsec\ radius  around the nebular
core, was detected in this previous data. This detection is illustrated
in the greyscale representation of the whole EW longslit spectrum 
shown in Fig. 7. and the relative surface brightness profile of this
\heiil\ emission (see Table 1) compared with 
that of the \niil\ emission along the same EW slit lengths 
is shown in Fig. 6. The \heii\ brightness profile from O'Dell et al (2004)
and that shown here for \heiil\ in Fig. 6 suggests that the He{\sc ii}
emitting region is a roughly  spherical volume.

Each \heiil\ profile was extracted for the incremental lengths along
the EW slit lengths listed in column 1 of Table 1. Each profile was
simulated by a least squares best fit single Gaussian profile whose observed 
centroid, corrected for heliocentric motion (1994 20 Sept.) 
is given in column 2 and observed halfwidth in column 3 of Table 1. 
These observed profiles are intrinsically broadened by the 19 
fine structural components spread between 6559.769--6560.209\AA\
which is $\equiv$~20~\kms\ range (see the `Atomic Line List' in
http://www.pa.uky.edu/~peter/atomic/). 

 \begin{table*}
\caption{Measured parameters of the \heiil\ profiles along the EW slit 
lengths.}
\label{12tab}

\begin{tabular}{ccccc}
\hline
1 & 2 & 3 & 4 & 5 \\
  extracted length    & observed central wavelength   &  
observed halfwidth & corrected halfwidth & profile brightness   \\
  arcsec     &  \AA\ (heliocentric)  & \AA\ & \kms\ & relative units 
(per increment)\\
\hline

     -64 to -27 & 6559.49 $\pm$ 0.06 & 0.61 $\pm$ 0.10 & 23 $\pm$ 4 & 0.90 \\
     -25 to -5  & 6559.50 $\pm$ 0.06 & 0.68 $\pm$ 0.05 & 27 $\pm$ 2 & 0.94 \\
      10 to 35  & 6559.54 $\pm$ 0.04 & 0.68 $\pm$ 0.05 & 24 $\pm$ 2 & 1.00 \\
      35 to 60  & 6559.53 $\pm$ 0.05 & 0.61 $\pm$ 0.15 & 23 $\pm$ 5 & 0.96 \\
      60 to 85  & 6559.53 $\pm$ 0.03 & 0.62 $\pm$ 0.08 & 24 $\pm$ 3 & 0.84 \\
      85 to 110 & 6559.53 $\pm$ 0.03 & 0.62 $\pm$ 0.04 & 24 $\pm$ 1 & 0.64 \\
     110 to 160 & 6559.54 $\pm$ 0.05 & 0.55 $\pm$ 0.04 & 20 $\pm$ 1 & 0.38 \\
     160 to 180 &   -      &     -            &  -  & 0.19 \\
     180 to 200 &   -      &     -            &  -  & 0.11 \\ 
\hline
\end{tabular}
\end{table*}

Following Clegg et al (1999), who calculated the relative brightnesses
for the fine structural components of other He{\sc ii} lines, 
the relative brightnesses of the 19 components for \heiil, 
within Menzel's Case B,
for electron densities of n$_{e}$ = 100 and 1000 cm$^{-3}$ were derived and 
are listed in columns 2 and 3 of Table 2 respectively. This density range 
is thought to be most appropriate for the \heiil\ emitting volume of
NGC~7293 and comparison between the values in columns 2 and 3 indicates
that the relative brightnesses have only a small dependence on  n$_{e}$
at these relatively low densities. The departure coefficients
from LTE by Case B derived by Storey \& Hummer (1995) were used.

\begin{table}
\centering
\caption{Relative strengths of the \heiil\ line 
fine--structure components normalized to the brightest component 
(6560.1848 \AA).}
\label{table1}
\begin{tabular}{c|c|c}
\hline
Fine structure components &
\multicolumn{2}{|c|}{Relative Strengths (CASE B, T $= 10^{4}$ K)} \\
of He{\sc ii} 6560.1 \AA\ (6 $\rightarrow$ 4) &  n $= 10^{2}$ cm$^{-3}$ & 
n $= 10^{3}$ cm$^{-3}$ \\
\hline
6559.7687 & 0.0535 & 0.0540 \\
6559.7940 & 0.0532 & 0.0536 \\
6559.8544 & 0.0184 & 0.0185 \\
6569.8874 & 0.0266 & 0.0268 \\
6560.0523 & 0.2983 & 0.3000 \\
6560.0528 & 0.0963 & 0.0971 \\
6560.0832 & 0.0011 & 0.0011 \\
6560.0839 & 0.0107 & 0.0108 \\
6560.1416 & 0.7709 & 0.7709 \\
6560.1418 & 0.4268 & 0.4293 \\
6560.1571 & 0.0001 & 0.0001 \\
6560.1573 & 0.0213 & 0.0214 \\
6560.1696 & 0.0368 & 0.0370 \\
6560.1766 & 0.0056 & 0.0057 \\
6560.1848 & 1.0000 & 1.0000 \\
6560.1882 & 0.0016 & 0.0016 \\
6560.1883 & 0.0101 & 0.0102 \\
6560.1941 & 0.0285 & 0.0285 \\
6560.2096 & 0.0023 & 0.0023 \\ 
\hline
\end{tabular}
\end{table}

Five fine structural components in Table~2, separated by 0.1326 \AA\ ($\equiv$
6 \kms) are dominant (i.e. 6560.0523, .0528, .1416, .1418 \& .1849)
therefore for the present purposes it is reasonable to simulate the effective
fine structural broadening by a Gaussian of this width.
 
The \heiil\ profiles are further broadened by a) the thermal motion of the
10$^{4}$~K gas (Gaussian halfwidth = 10.7~\kms) b) the effective instrumental
halfwidth of 11 \kms\ (from measurements of the lamp lines) c) bulk
turbulence in the emitting volume and d) radial expansion of the same volume
which is the parameter of particular interest.
The corrected halfwidths in column 4 of Table 1 are the observed halfwidths
in column 3 corrected for both thermal and instrumental broadenings,
for these are well--known, 
assuming that all broadening functions are Gaussian. Finally the relative 
\heiil\ profile brightnesses in column 5 are the mean values per increment
for the whole of the line profiles along the incremental lengths in column 1.
The continuum emission adjacent to the \heiil\ line (see Fig. 7) 
was subtracted in this process.
The profiles from the last two incremental lengths listed in Table~1
were so faint that only their relative surface brightnesses could be
measured with significant accuracy.

\ha, \niil\ and \heiil\  profiles extracted from a sample 
length of 60
arcsec centred to the east of the central star (see Fig. 7) 
are shown in Fig. 8.
This range just excluded contamination by the stellar spectrum
and gave the highest signal to noise ratio for the critical 
but faint \heiil\ line.
A similar detection of \heiil\ emission was made in the NS spectrum 
(not shown).

The core of the central spherical volume of \heiil\ emission  
is surrounded, and partially coincident with, 
an inner 
shell of \oiiil\ emission whose emission peaks are separated
by $\approx$ 240 arcsec (Meaburn \& White 1982 and see fig. 4 
in O'Dell et al 2004). 
This shell can be seen clearly in the
deep negative representation of the \oiiil\ image taken with
the New Technology Telescope (La Silla) in Fig. 9
(see Meaburn et al 1998 for the technical details). Spatially
resolved \oiiil\ line profiles were obtained with MES
combined with the AAT along the 163 arcsec long E--W slit
marked in Fig. 9. The integration time was 1800 s and slit width
70 $\mu$m ($\equiv$ 6 \kms\ and 0.5 arcsec on the sky). Again
the technical details of this as yet unpublished 
observation are given in Meaburn et al (1998).
Co--added profiles were obtained for all blocks, 5 increments long, along the
512 CCD increments that covered the slit length. The separate
velocity components in each of these profiles were simulated
by Gaussians whose centroids are shown in Fig. 9. Over the central
star (at 0 arcsec offset in Fig. 9) it can be seen that four velocity
components in the \oiiil\ profiles are present 
(i.e. at \vhel\ = $-$60, $-$42, 
$-$17 and $-$4 \kms). The \oiiil\ line profile
for an incremental length comparable to that used
for the \ha, \niil\ and \heiil\ profiles is also shown in Fig. 8. 

\subsection{Imagery of radial `spokes'}

The cometary knots have dense, molecular heads (Meaburn et al 1992 and
Huggins et al 1992) with tails extending away from the central
stars. O'Dell et al (fig. 16 -- 2004) and Henry et al (fig. 5 --- 1999) 
have revealed
even more widespread system of 
headless `spokes' in the \nii\ emitting gas extending
from the inner torus outwards in the bright nebulosity. Henry et al (1999)
showed them as parallel ridges of enhanced \nii\ emission by dividing their
\nii\ image with their \ha\ one.

An image of 1800 s duration through a 72~\AA\ bandwidth
filter centred on  \ha\ + \nii\ was obtained with the NTT
(see Walsh \& Meaburn 1993; Meaburn et al 1998 for the technical details) 
and is shown
at high contrast in Fig. 10. The outer  radial `spokes'
can be seen to continue faintly inside the disk containing the
cometary globules to within 
$\approx$~30\arcsec\ from the central
star.
They
are seen to be 
radially distributed with respect to the central star and continuous
in some cases with the more recently discovered, but similar features
in the adajcent bright helical structure (O'Dell et al 2004; Henry et al 1999).
For these reasons  
they are considered to be real and shown here
at high contrast. Obviously a deep, reasonably high resolution, image
in the same emission lines is needed to confirm their presence
for the image in Fig. 10 is not flat--fielded. However, these spokes
are relevant to the understanding of cometary globule formation
and knot evaporation.
 
\section{Kinematics and morphologies}
\subsection{\heiil\ volume and \oiiil\ inner shell}

The presence of the \heiil\ volume close to the central star
affects many of the theoretical ideas for the creation of planetary
nebulae consequently its structure and motions should be quantified.
The \heiis\ image of O'Dell et al (2004) suggest that it is
a spherical volume of $\approx$~180\arcsec radius. This morphology
is confirmed by the relative brightness distribution here in Fig.6.
The widths of the \heiil\ line profiles when corrected for thermal
and instrumental broadenings (column 4 of Table 1) have a maximum width
of 26 $\pm$ 2 \kms\ near the central star with only a minor (and  uncertain) 
decline in width up to 160 arcsec from this star. As this width will
also contain the consequences of fine structural width and 
turbulent motions then the 13 $\pm$ 2 \kms\
is very much an upper limit to the radial expansion of the \heiil\ emitting
volume. If turbulent motions of  $\geq$ 10 \kms\ (flows at
the sound speed off globule surfaces for instance)
contribute to the line width  in addition to a
 fine structural broadening component of $\approx$~6~\kms\
 then a global expansion of the \heiil\ emitting region
would be $\leq$~12~\kms. The non--Gaussian shape of the \heiil\
profile in Fig. 8 could suggest that an expansion velocity near
this upper limit is in fact present.

The inner \oiiil\ emitting shell in Fig. 9 is co-existent with the
outer, and faint, parts of the \heiil\ emitting volume.
The splitting of 25 \kms\ of the \oiiil\ profile 
over the central star, which is symmetrical around \vsys,
undoubtably is caused by radial expansion of 12.5 $\pm$ 1.0 \kms\ of 
the inner \oiiil\ emitting shell. This value is now considered to be
more realistic than the $\approx$ 20 \kms\ given in Meaburn et al (1998)
derived from Fabry--Perot observations in Meaburn \& White (1982 -
one velocity component was misinterpreted).

Incidentally, because this \heiil\ emitting region is relatively inert and 
so close to the central star, and because
it is a single recombination line with a well known wavelength
though complicated by fine structural components,
a bi--product of the present
\heiil\ profile measurements
is a reliable value of the systemic heliocentric radial
velocity (\vsys) for NGC 7293.As the mean observed
central wavelength is 6560.053   
$\pm$ 0.015 \AA\ (mean of values in column 2 of Table 1) 
and with the mean rest wavelength, with each fine structural
component weighted by its relative brightness in Table 2, is
6560.128 \AA\ 
and as \vhel\ = \vobs\ $-$24.3 \kms\ on the date of observations 
then \vsys\ = $-$27.1~$\pm$~2.0
\kms\ when systematic calibration uncertainties are included.
This is consistent with a previous, though less certain,
value of \vsys\ derived from the centroids
of the split \niil\ profiles over the central star (Meaburn et al 1998):
the \niil\ components are not necessarily symmetric around \vsys.

\subsection{The bright helical structure}

The bright helical structure emitting \niil\ can be seen in 
Meaburn \& White (1982) and O'Dell et al (2004) to be enclosing
an `outer' \oiiil\ emitting shell, the `inner' \oiiil\ shell shown in
Fig. 9 as well as the central \heiil\ emitting volume (see Fig. 6).
It was proposed in Meaburn et al (1998) that this helical structure could
be a consequence of bi-polar lobes expanding from
a toroidal ring, itself expanding radially. As O'Dell et al (2004) challenged
this assertion the kinematics and morphology of the \niil\
helical structure has been further explored using the XSHAPE code
developed by Steffen, Holloway \& Pedlar (1996)
and Harman (2001). This numerical technique does not give a unique
answer but does predict accurately the observed surface brightnesses and
spatially resolved
radial velocity shifts of emission lines for any predetermined structure.

In the present case a starting point for this modelling is to 
identify as B and C the 
manifestation
of the central toroid in the E--W pv array in Fig. 6. These positions
coincide with major brightness maxima and have radial velocities
nearly symmetrically displaced from \vsys\ = $-$27 \kms. Undoubtably
the ionised inside surface of a larger, radially expanding,
molecular toroid (Healey \& Huggins 1990) 
is being observed at an angle 
in agreement with O'Dell et al (2004). In the very
simplest of bi--polar models the positions A and D in Fig. 6 are where
these sight lines intersect the edges of the lobes. As the observed 
radial velocities
at both of these positions are close to \vsys\ this supposition is
reasonable if it is assumed that lobe expansion is always
proportional to the distance from the central star and along a vector
pointing away from this centre.

The model shown in Fig. 11 is based on these assumptions and using the
XSHAPE code the predicted morphology is compared to the observed image 
from Fig. 1 and the negative greyscale representations of both the 
observed N--S
and E--W pv arrays of \niil\ profiles from Meaburn et al (1998).
The positions A -- D marked in Fig. 6 are also shown.

In this model
the tilt of the sightline to the nebular axis is 37\degr, the
radial expansion of the toroid (B \& C) is 14.25~\kms\ and the 
expansion of the
two lobes 24.5~\kms\ along the bipolar axis. Each lobe has the
shape of an inverse ellipsoid (sphere divided by an ellipse) with semi--major 
axis of 490\arcsec\ and semi-minor 285\arcsec. 
To give the appearance of reality,
clumps are distributed randomly within the 3D grid with
0.0001 clumps per cubic arcsec. Each clump has the form of a 3D
Gaussian with FWHM  of 20\arcsec. The simulated image has been
smoothed with a 10\arcsec\ wide Gaussian to match the resolution in Fig. 1.
The simulated pv arrays have been convolved with the 11~\kms\ wide 
instrumental profile.
A convincing, though obviously not perfect, fit of the model
predictons to the salient features of the observed
image and the two orthogonal pv arrays is demonstrated 
in Fig. 11. These observations and model cannot accomodate
the presence of two tilted rings as suggested by  O'Dell et al (2004).

The model in Fig. 11 is only for the bright \niiab\ emitting helical
structure which has its CO emitting inner counterpart (Young et al 1999).
O'Dell et al (2004) suggest that the inner ring (whose
edges are marked as B and C in Fig. 11) is an expanding disk 
as modelled here. However, they go on to suggest that the two
adjacent bright arcs (whose edges are at A and D respectively in Fig. 11) 
are the manifestation
of a single expanding disk expanding {\it orthogonally} to
the central disk. They base this assertion soley on fig. 5
of Young et al (1999) where an orthogonal range of velocities
for a CO `outer arc' is partially detected. The excellent 
morphological and kinematical fit of the model in Fig. 11 (and see
fig. 9 of Young et al 1999) suggests that this interpretation is incorrect.
An outer partial CO ring, expanding othogonally but faint in \niiab\ (Young
et al 1999) must exist but is not part of the bright \niiab\ emitting 
helical structure
modelled here.

\subsection{The halo}

The nebulous N--E halo arc imaged by Malin (1982) and see also
Speck et al (2002) 
and shown
in Fig. 1b and  Fig. 2 has a S-W counterpart which could 
suggest some form
of bi--polar
structure which is both larger and along a different axis to that proposed
for the bright helical nebulosity. However, the pv arrays of \niil\ profiles
in Figs. 3--5 along slits 1--3 in Fig. 2, and those of Walsh \& Meaburn
(1987) reveal behaviour not easily explained in a model
involving classical three dimensional bi--polar lobes.
For instance, although there is a systematic trend in radial velocity
from the north of slit 2 (\vhel\ = 0 \kms) to the southern end 
where \vhel\ = $-$60 \kms, which suggests some form of expansion, 
the bright northern edge would be expected at \vsys\ = $-$26 \kms\ in a simple
bi--polar model.The eastern edge itself was shown to have complex
velocity structure with an extreme velocity  component at 
\vhel~=~$-$80 \kms\ (slit 3 in Fig. 5 and also see
Walsh \& Meaburn 1987).
Nevertheless, line spitting observed in slit positions 
2 \& 3 (Figs. 4 \& 5) suggests the interpretation of
these features as expanding lobes though not orthogonal
to the central torus which is modelled in Fig. 11. A similar
case is observed in NGC 2440 (L{\'o}pez et al (1998).

An alternative possibility is that both this eastern and 
western halo structure
could rather be both parts of a single expanding ring of ionised material. 
If so it must be oriented at a completely different angle to the central
torus (BC) shown in Fig. 11 for the eastern edge has predominantly
approaching radial velocities with respect to \vsys. If these 
reflect the radial expansion of the ring then its expansion
could be as high as $\approx$~70 \kms\ for an arbitary
tilt of 45\degr\ to the sight line although the observed radial 
velocities do not unambiguously favour this ring model.

More observations at similar spectral resolution, particularly
of the south--western halo lobe, are required to investigate
these propositions more fully.

Other features of the general halo of NGC 7293 have been
seen previously (e.g. O'Dell et al 2004) but of potential interest 
is the faint jet--like feature in the SW, along PA 255\degr, but projecting
back to the nebular core in the deep
image in Fig.~1c. A bow--shaped possible counter feature on the opposite 
side of the nebula is aligned with it. 

\section{Discussion}

\subsection{The interacting winds model}

PNe are a consequence of the sequence of events as a star of $\leq$ 8 \msun\
evolves through its Red Giant (RG) and  Asymptotic Giant Branch (AGB) 
phases to finally become a White Dwarf (WD). In this sequence it is
generally accepted that an initial  RG wind, is followed
by the more volatile and many times denser 
AGB `superwind' (with sporadic outbursts at 20 of \kms)
which is then subjected to interaction with a 
high--speed wind (several 1000 \kms)
and becomes photo--ionised as the central star becomes a 
WD with a high surface
temperature. At this point the circumstellar envelope becomes the 
embryo PN.
The fast wind could then decline as the star becomes an older WD and the
photoionised circumstellar 
envelope would then be described as an evolved PN.

The standard interacting winds (IW) model (Kwok, Purton \& Fitzgerald 
1978; Khan 
and West 1985; Chu, Kwitter \& Kaler 1993; Balick \& Frank 2002) makes 
a theoretical attempt
to 
explain the formation of the observed nebular structure 
and kinematics in PNe 
during the transition of the stellar core from the post-AGB to 
the WD stages. Here, with smooth, isotropic, density distributions
the fast wind forms an expanding pressure driven (energy conserving)
bubble in the pre--cursor AGB wind. A central volume of unshocked
fast wind is surrounded by a shell of superheated (10$^{6}$ K) shocked
wind which provides the gas pressure to form and 
accelerate a shell of shocked
AGB wind. Within this model and with photo--ionisation dominant 
the latter expanding, low--ionisation, shell is characteristic
of young PNe.  Balick \& Frank (2002) emphasise that
from observations of a large number of PNe this model, although
an excellent theoretical starting point, is very
idealistic; there are a multitude of phenomena within PNe when the
stars are in their post--AGB phases
that require more complex explanations; lobes and disks expanding along 
separate axes being only two of many such complexities.
 
The present kinematical observations of  the undoubtably 
evolved PN, NGC 7293, reported here
could be consistent with the final stages of the IW model if it
is assumed that the fast wind has now switched off. This possibility 
is made 
plausible by the absence of a direct observation of the fast wind in the
stellar spectrum (see Sect. 1) and the very presence of 
an inner \heiil\ volume
and \oiiil\ shell would prevent the direct interaction of such a wind
with the outer helical structure. The progression outwards of the expansion
velocities i.e.  $\leq$~12~\kms\ for the \heiil\ emitting
centre, 13 \kms\ for the surrounding inner  \oiiil\ shell
and 25 \kms\ for the \niil\ emitting outer envelope albeit inside
an outer bi--polar configuration (Fig. 11), could be a consequence 
of the sudden pressure
decline 
that would occur 
as the fast wind switched off leading to acceleration 
inwards of the now unsupported inner regions of the previous pressure 
driven shell (private comms. by Steffen and Dyson).
Incidentally, Wilson (1950) had noticed this increasing velocity
from the high--ionisation centres to low--ionisation envelopes of
many similarly evolved PNe. 

However, the clumpy nature of most of the circumstellar material also
poses a direct problem for the IW model (see Sect. 4.2).

\subsection{Cometary knots and \niil\ `spokes'}

The radial `spokes' that are prominent in the bright helical
nebulosity and shown here to extend, faintly, to within 30\arcsec\ 
of the central 
star are most likely a consequence of the clumpy nature of the nebula. 

Such a clumpy structure is to be expected when the whole of NGC 7293 
is considered for it is famous for its cometary globules
beyond a radus of 90\arcsec\ from the central star.
These were first shown to have dense molecular cores 
directly by Huggins et al (1992) and by Meaburn
et al (1992) when sillouetted against the central \oiiil\ emission. 
Dyson et al (1989) had suggested that the cometary globules
were a consequence of the ejection of the RG maser spots as dense
globules
later overrun by the faster AGB wind. If this clumpiness extends
to the nebular core it would suggest that clump ejection had occurred
with a wide spread of ejection velocities. The system of well-known
cometary globules is expanding radially within the central
torus at 14~\kms\ consistent with
their RG origin (Meaburn et al 1998).

The headless radial `spokes' could then all be relics
of cometary tails from globules that have nearly photoevaporated.
The inner `spokes' in Fig. 10
could be  elongated enhancements of \niil\ emission either
in the torus containing the cometary globules, extending
inwards, or in the walls of the 
lobes in the model
in Fig. 11 but pointing always towards the central star. In the 
latter case they are
then seen in projection over the inner \oiiil\ and \heiil\ emitting
core of the nebula.

The recombination time T of hydrogen is $\approx$~10$^{5}$ n$_{e}$$^{-1}$ yr
then for a tail electron density n$_{e}$ = 50 cm$^{-3}$, T $\approx$
5000 yr.
A longer recombination time in the lower density ionised material 
of a tail, after the dense head of a cometary globule has evaporated, 
could on its own then explain the tail's  persistence.
A more detailed explanation could be given by
L{\'o}pez-Martin et al (2001) who have discussed the evaporation 
of the  
cometary globules  and their 
tails (see also Cant{\'o} et al 1998). The clump heads are 
exposed to the direct ionising radiation of the central star while the 
tails are only exposed to the diffuse radiation field. This is 
typically a few percent of the direct field at the location of the 
bright optical knots but is very much less closer to the star because 
of the $r^{-2}$ rise in the direct flux (L{\'o}pez-Martin et al 2001). For 
clumps that are close to the central star, recombinations in the 
evaporating head gas can not significantly absorb the stellar flux. In 
this case, the evaporation timescale is approximately (L{\'o}pez-Martin et 
al 2001)
\begin{equation}
t_{\rm evap}\simeq \frac{R_{\rm c}c_{\rm i}}{c_{\rm n}^2}
\end{equation}
where c$_{\rm i}$ and c$_{\rm n}$ are the sound speeds in the ionised
and neutral gas respectively. 
Note especially the dependency in Eq. 1 
on clump radius R$_{\rm c}$: once a clump begins to 
shrink the evaporation timescale drops and the clump will then rapidly 
disappear. The tail will only begin to be exposed to the direct stellar 
radiation field when the clump shrinks appreciably. It is thus entirely 
plausible that the tail may survive the destruction of its parent 
clump. It will then be quickly ionised by the direct field and be very 
overpressured but as is clearly seen in Fig. 10 the clumpless tails 
have not yet had time to mix with the surrounding gas (for a tail of 
radius $10^{15}$ cm the dynamical timescale is a few thousand years in 
the ionized gas).

This predominance of 
clumps
in the ionised envelope of NGC 7293 also makes it most likely
that momentum--conserving, mass--loaded (ablated) flows (Dyson, Hartquist
\& Biro 1993)
have played an important, even dominant, part in the evolution
of this PN. A hybrid explanation, with the IW model applicable immediately
after the onset of the fast wind followed by a domination of ablated flows 
up to the switch--off of the fast wind, may be a more appropriate way
of considering the creation of NGC 7293.The velocity structure
of the clumpy PNe such as NGC 7293 has been recently studied by
Steffen \& L{\'o}pez (2004) where progressive expansion velocities
with distance from the central star are obtained in the simulations.   

\subsection{The halo features}

The extended ionised material beyond the bright inner regions (Fig. 1)
was 
detected by Speck et al (2002) using
Southern H--Alpha Sky Survey Atlas data (SHASSA--Gaustad et al
2001). Meanwhile, O'Dell et 
al (2002) carried out detailed photoionisation modelling of the bulk of 
the inner shell using {\sc cloudy}. The modelling indicated that the 
nebula was ionisation bounded rather than density bounded yet the 
presence of the extended ionised gas is confirmed here and in O'Dell et 
al (2004). The apparent contradiction between the {\sc cloudy} results 
of O'Dell et al (2002) may be explained in the following qualitative 
way. The very clumpy nature of the inner ring of NGC 7293 (the 
majority of the mass of the shell may even reside as molecules 
in the clumps-- Speck et al 2002) means 
that it is possible, as long as the volume filling factor of the clumps 
is not too great, for the ionising radiation to escape the inner ring. 
The photoionisation modelling will yield results consistent with an 
ionisation bounded system simply because the bulk of the emission comes 
from the clump surfaces where the ionisation is locally bounded. 
Alternatively, O'Dell et al (2004) maintain that the central
torus (BC in Fig. 11) is globally 
ionisation bounded and that the extended ionised gas lies above and 
below the  plane of this torous. 
They also suggest that the extended ionised gas 
could have been ionised at an earlier epoch and owing to a low density 
has not yet recombined.

The possibility that expanding lobes or even a disc of material (Sect. 3.3)
exists in the
halo of NGC 7293 is not unexpected when other PNe are considered. Multiple 
axes of ejection are observed in the PNe NGC 6302 (Meaburn \& Walsh 1980), 
KjPn8 (L{\'o}pez et al 2000), NGC 2440
(L{\'o}pez et al 1998) and J 320 (Harman et al 2004) for example. A similar 
high--speed disc
is observed around the bi-polar lobes of Mz~3 by Santander--Garc{\'\i}a
 et al (2004).
These phenomena are most likely evidence of the volatility,
and as yet not understood behaviour,
of the AGB phase in the evolution of the central stars of PNe.

The possibility reported here of a jet and counter bow--shaped feature
in the halo of NGC 7293 is again not unexpected when the 
whole range of observed 
phenomena are considered in PNe. Collimated outflows are seen in many
PNe e.g. FG 1
(L{\'o}pez, Roth and Tapia 1993, L{\'o}pez, Meaburn \& Palmer 1993) etc. 
There is at yet no consensus about their origin
though they must be associated with the onset of the superwind as the
stars enter their post--AGB phase with binarity of the central
stellar systems playing some part in their generation.

\section{Conclusions}

Progressively increasing expansion velocities are observed 
from the highly ionised core of NGC 7293 to the lobes that constitute
the lowly ionised helical structure.

\noindent Numerical simulations of the latter structure as bi--polar
lobes, expanding perpendicularly to a central torus, give a reasonable match
to the observed images and pv arrays of line profiles.

\noindent The standard interacting winds model for the creation of 
PNe explains
this kinematical and morphological structure only if the fast wind
has switched off as the central star becomes a WD. Ablated flows
must add a further complication to any dynamical considerations. 

\noindent Radial `spokes' of \nii\ emission which are now traced to near
the central star could be the remnants of the cometary tails of
dense globules that have now photo-evaporated.

\noindent It is proposed that a prominant halo arc could be part of a single
expanding disk along a different axis to that of the bright nebula
although expanding bi--polar lobes along this same axis remain an
interpretation.  
More extensive kinematical observations are needed to distinguish
between these
possibilities.

\noindent A jet-like feature and its counter bow-shock are 
tentatively identified.

\noindent Multiple events in the AGB phase, and shortly
afterwards, of the central star must 
have created these phenomena.

\begin{figure*}
\epsfclipon
\centering
\mbox{\epsfysize=9in\epsfbox[20 20 330 822]{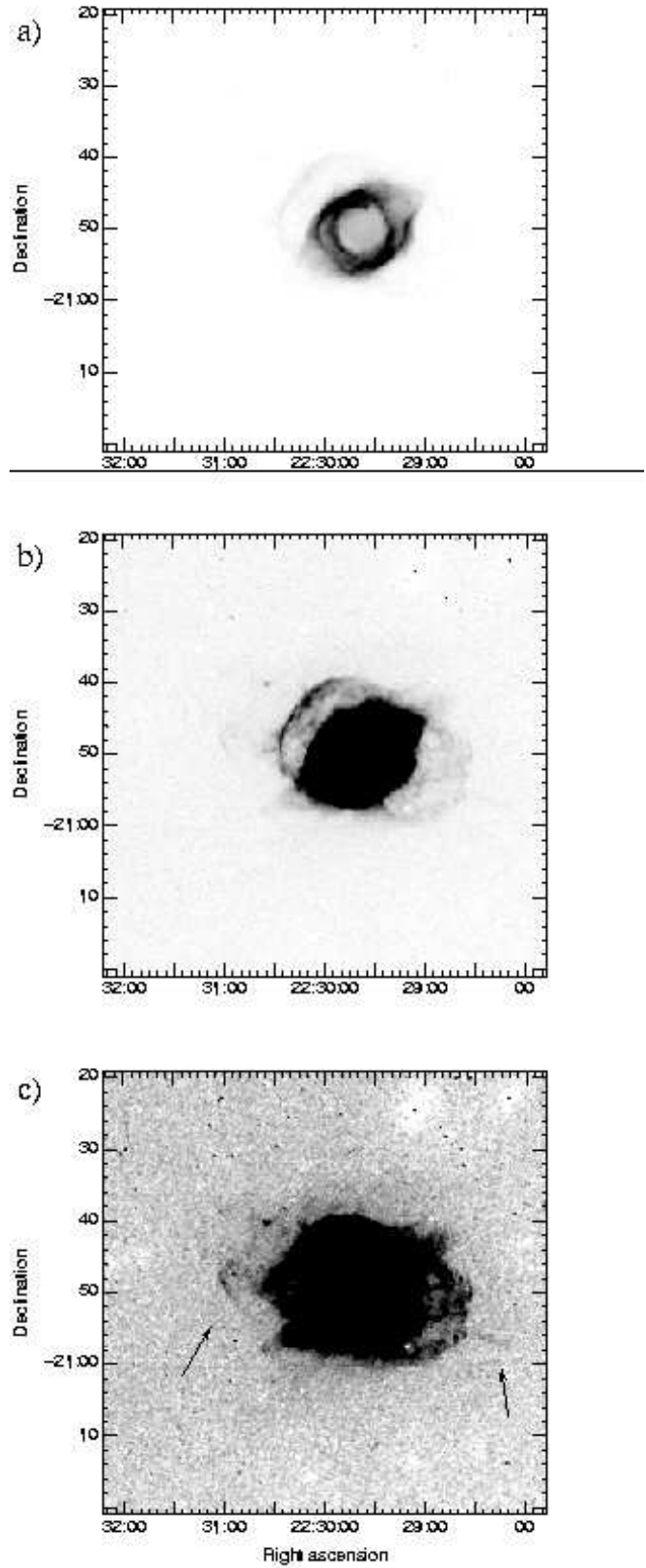}}
\caption{The same deep \hnii\ image is displayed at three 
contrast levels to show all of the features from the bright,
inner helical structure out to the very faint halo regions (epoch 2000).
The bow--shaped feature and jet--like feature are both arrowed in c). }
\label{reffig1}
\end{figure*}

\begin{figure*}
\epsfclipon
\centering
\mbox{\epsfxsize=6.5in\epsfbox[20 20 575 575]{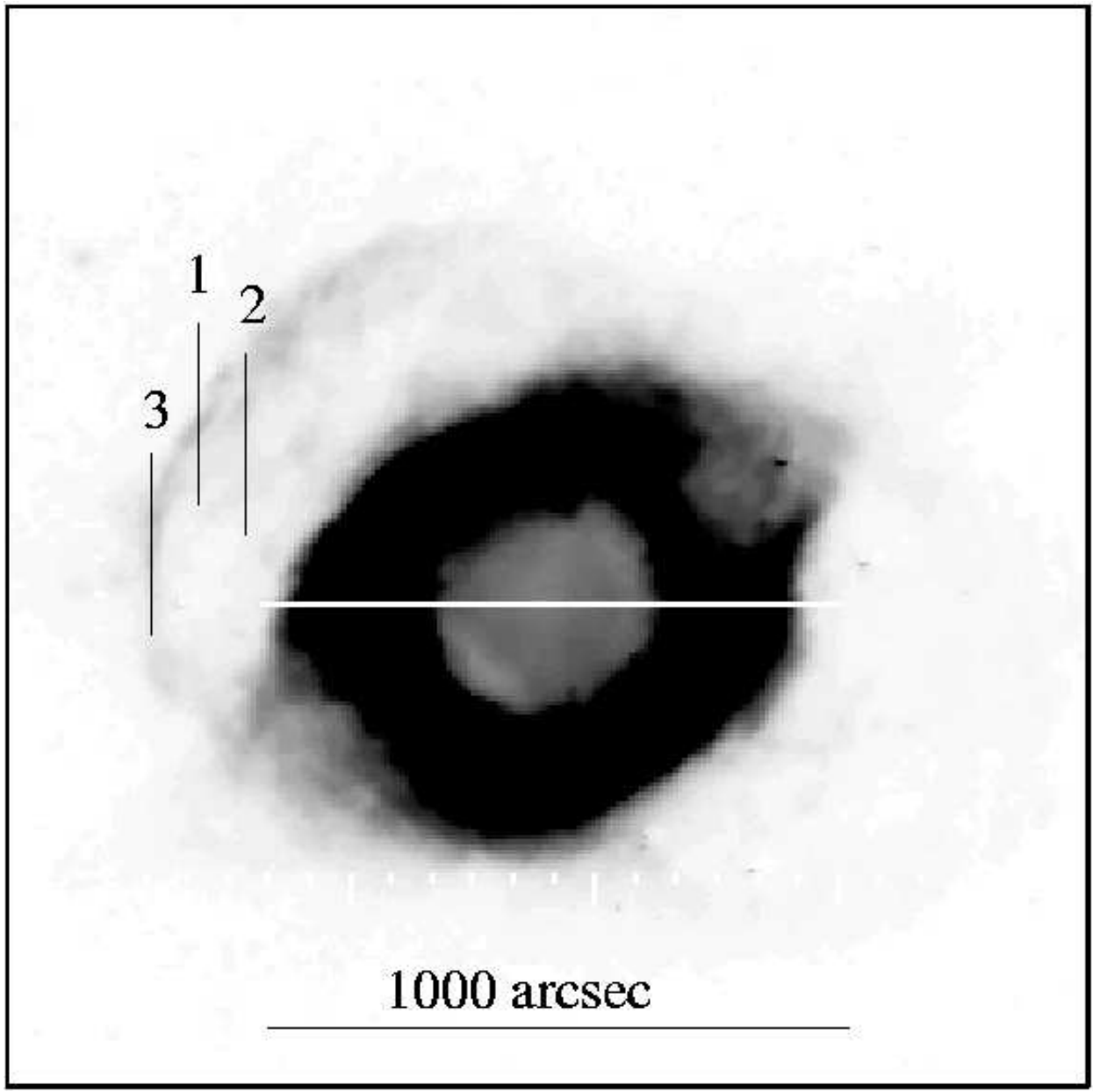}}
\caption{New slit positions (1--3) and the previous E--W continous
line of spectral measurements (Figs. 6 \& 7) are shown against
the \hnii\ image (Fig. 1).}
\label{reffig2}
\end{figure*}

\begin{figure*}
\epsfclipon
\centering
\mbox{\epsfysize=9in\epsfbox[20 20 364 822]{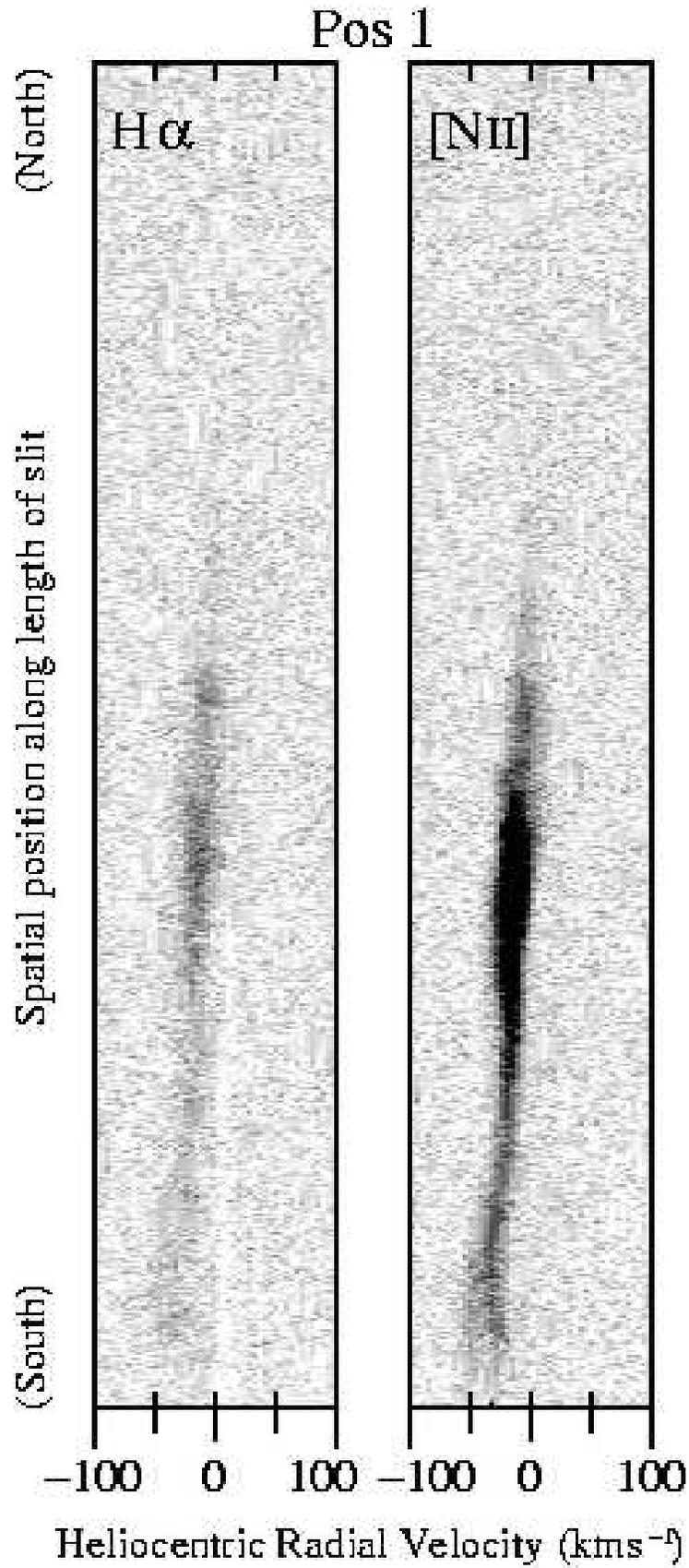}}
\caption{Greyscale representations of the \ha\ and \niil\ profiles
along slit position 1 in Fig. 2 are shown. }
\label{reffig3}
\end{figure*}

\begin{figure*}
\epsfclipon
\centering
\mbox{\epsfysize=9in\epsfbox[20 20 372 822]{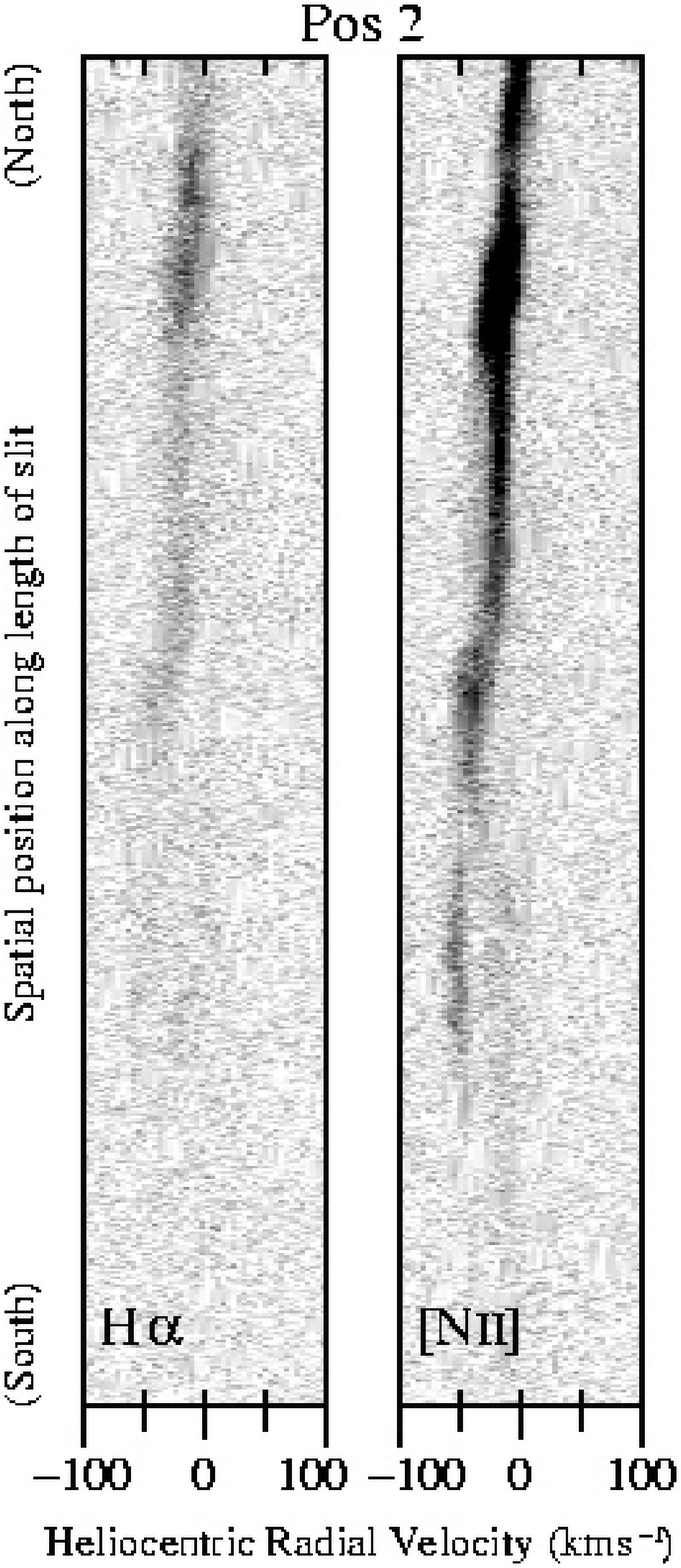}}
\caption{As for Fig. 3 but for slit 2}
\label{reffig4}
\end{figure*}

\begin{figure*}
\epsfclipon
\centering
\mbox{\epsfysize=9in\epsfbox[20 20 373 822]{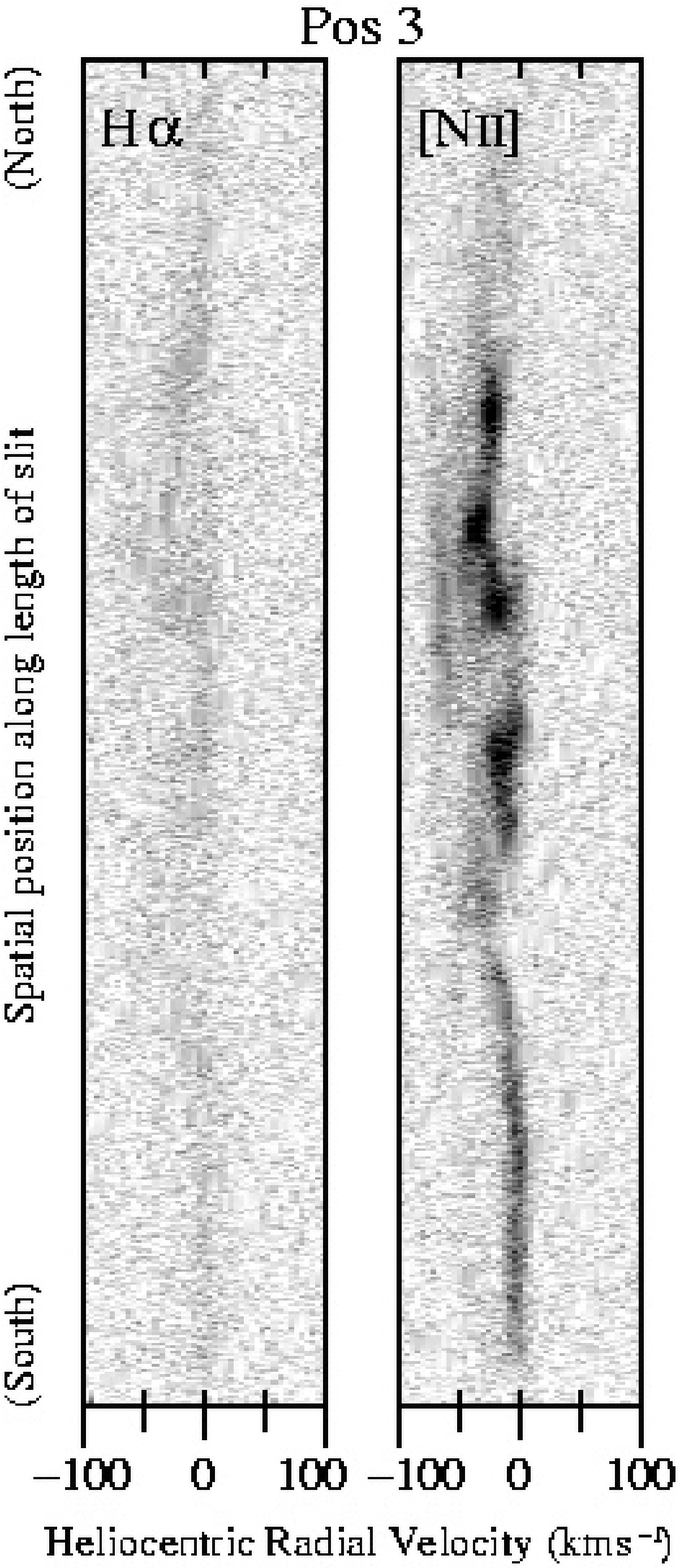}}
\caption{As for Fig. 3 but for slit 3.}
\label{reffig5}
\end{figure*}

\begin{figure*}
\epsfclipon
\centering
\mbox{\epsfxsize=7in\epsfbox[20 20 575 469]{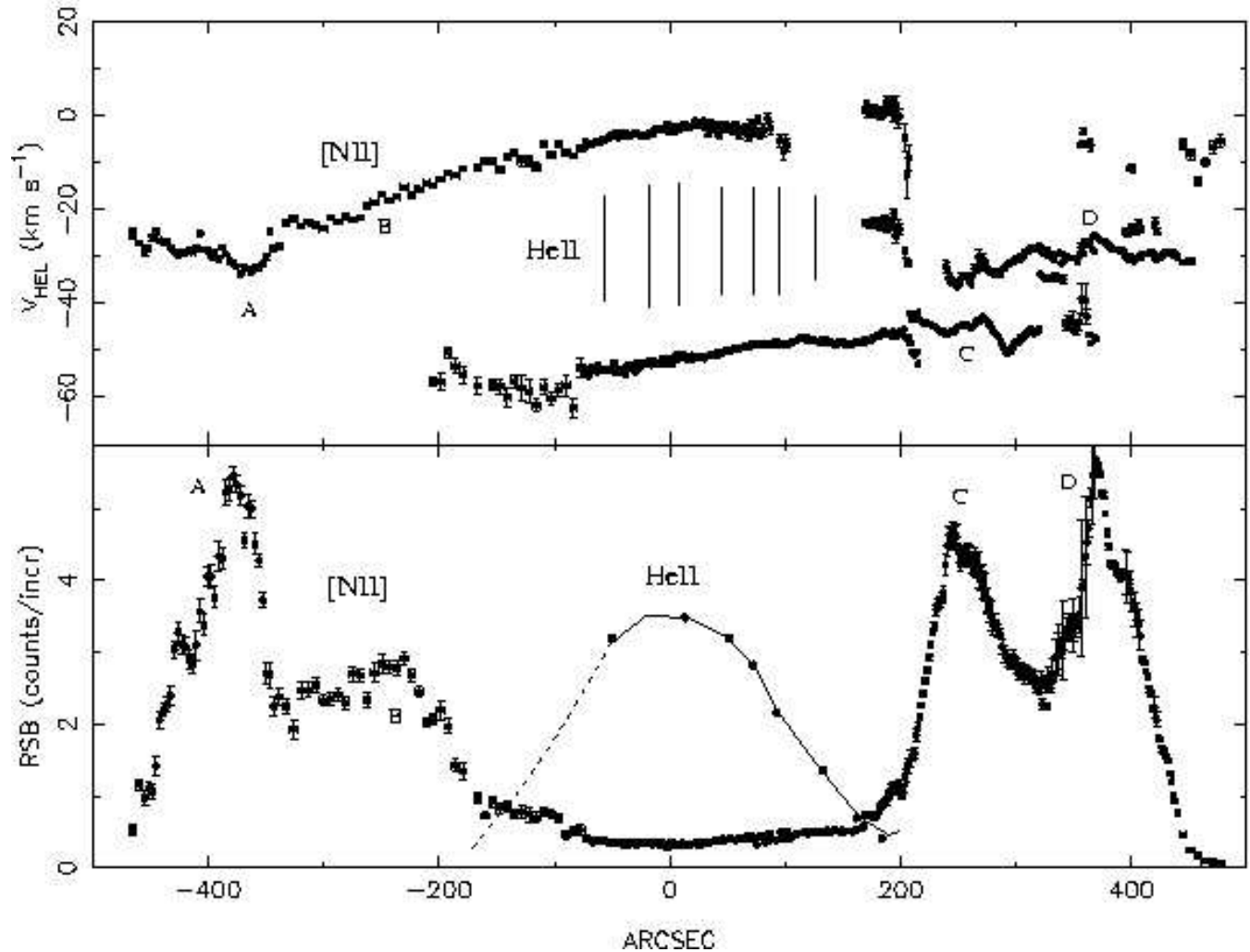}}
\caption{In the top panel the centroids of individual velocity
components in the \niil\ profiles, from the long E--W length marked in 
Fig. 2, are compared to the halfwidths of the \heiil\ line
profiles when corrected for both instrumental and thermal broadening
(column 4 in Table 1. In the bottom panel the relative surface brightnesses
(RSB) of the \niil\ and \heiil\ profiles are compared. The dashed
line is simply the western observed curve folded over to the eastern
side (where it was below the detection limit). This is justified
by the \heiil\ brightness variations observed along the NS slit (not shown).
Key positions are marked A--D in both panels.}
\label{reffig6}
\end{figure*}

\begin{figure*}
\epsfclipon
\centering
\mbox{\epsfxsize=7in\epsfbox[20 20 575 620]{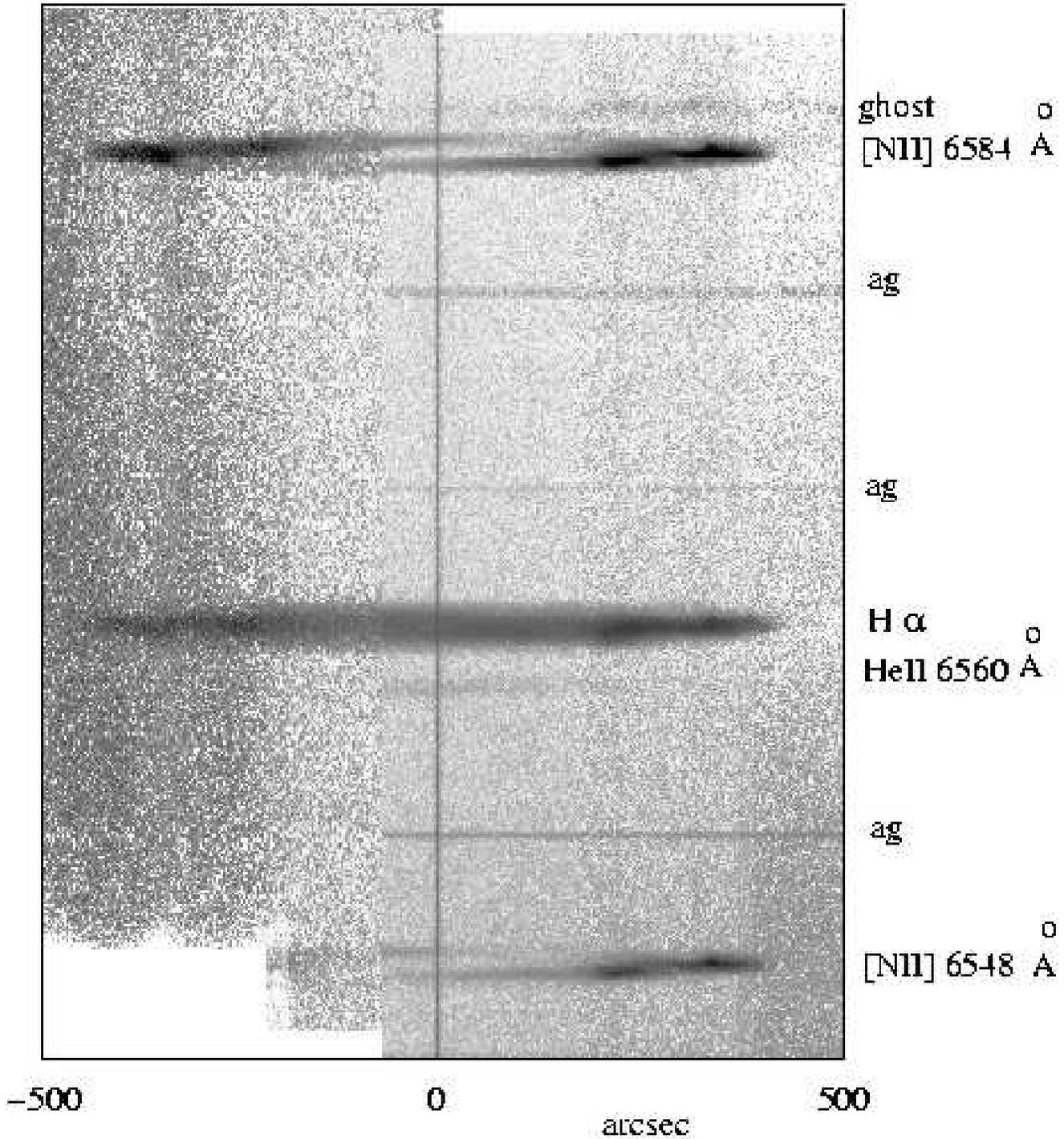}}
\caption {The negative greyscale 
representation of the combined spectra along the E--W length marked
in Fig. 2 is shown. The various nebular emission lines and airglow
lines (ag) are identified. The feature above the \niil\ profiles is an
optical `ghost'}
\label{reffig7}
\end{figure*}

\begin{figure*}
\epsfclipon
\centering
\mbox{\epsfysize=9in\epsfbox[0 0 355 1085]{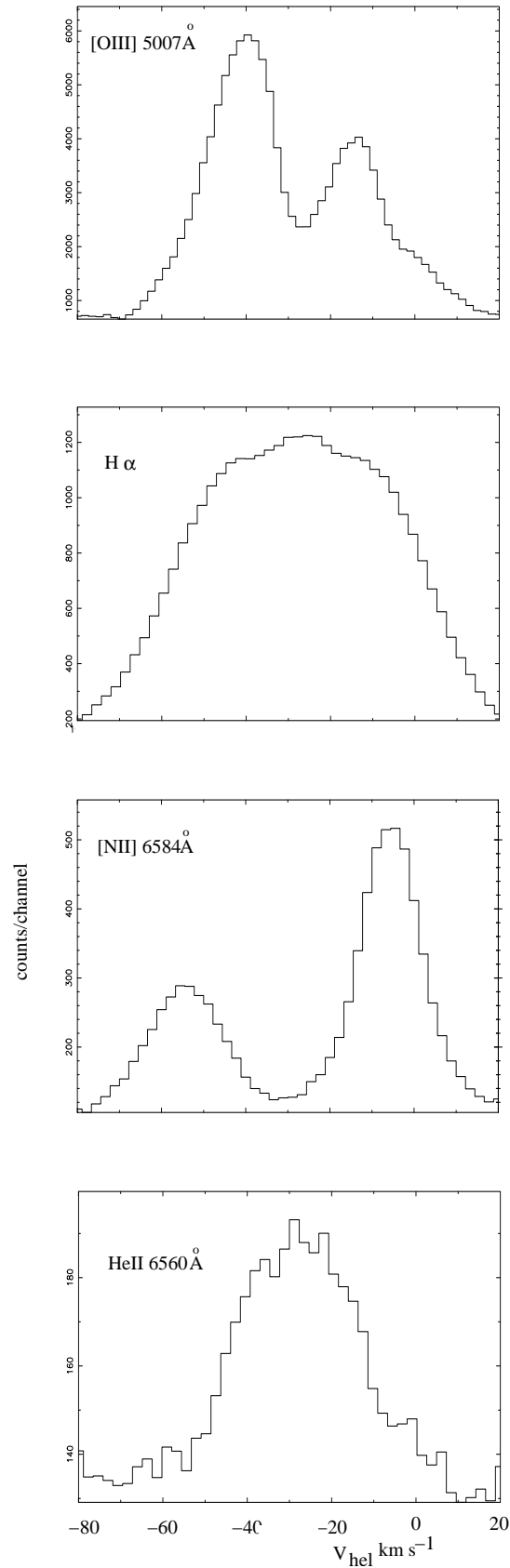}}
\caption{Various line  profiles from the same $\approx$ 60 arcsec
length of the E--W slit shown in Fig. 2 are shown. These extractions
were for this length centred 32 arcsec E of the central star to permit
a sufficient signal to noise ratio to be achieved for the faint
\heiil\ profile. If the stellar spectrum is included it is degraded
significantly. }
\label{reffig8}
\end{figure*}

\begin{figure*}
\epsfclipon
\centering
\mbox{\epsfysize=9in\epsfbox[20 20 550 822]{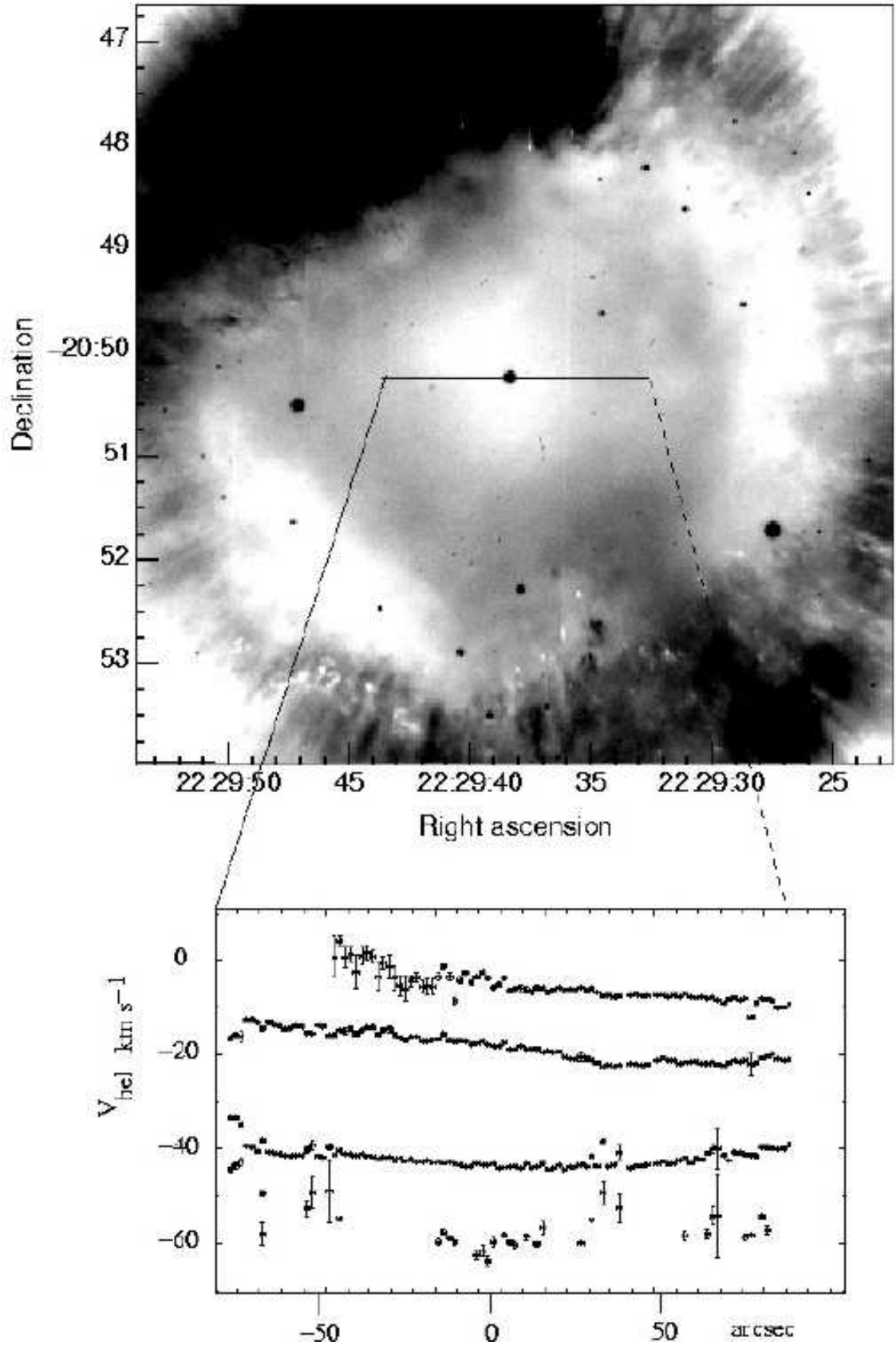}}
\caption{A deep NTT image in the light of \oiiil\ is compared
with centroids of the separate velocity components in profiles
of the same emission line along the slit position marked in 
the top panel. The
`inner' \oiiil\ emitting shell can be seen in the top panel (epoch
2000).}
\label{reffig9}
\end{figure*}

\begin{figure*}
\epsfclipon
\centering
\mbox{\epsfysize=6in\epsfbox[20 20 575 563]{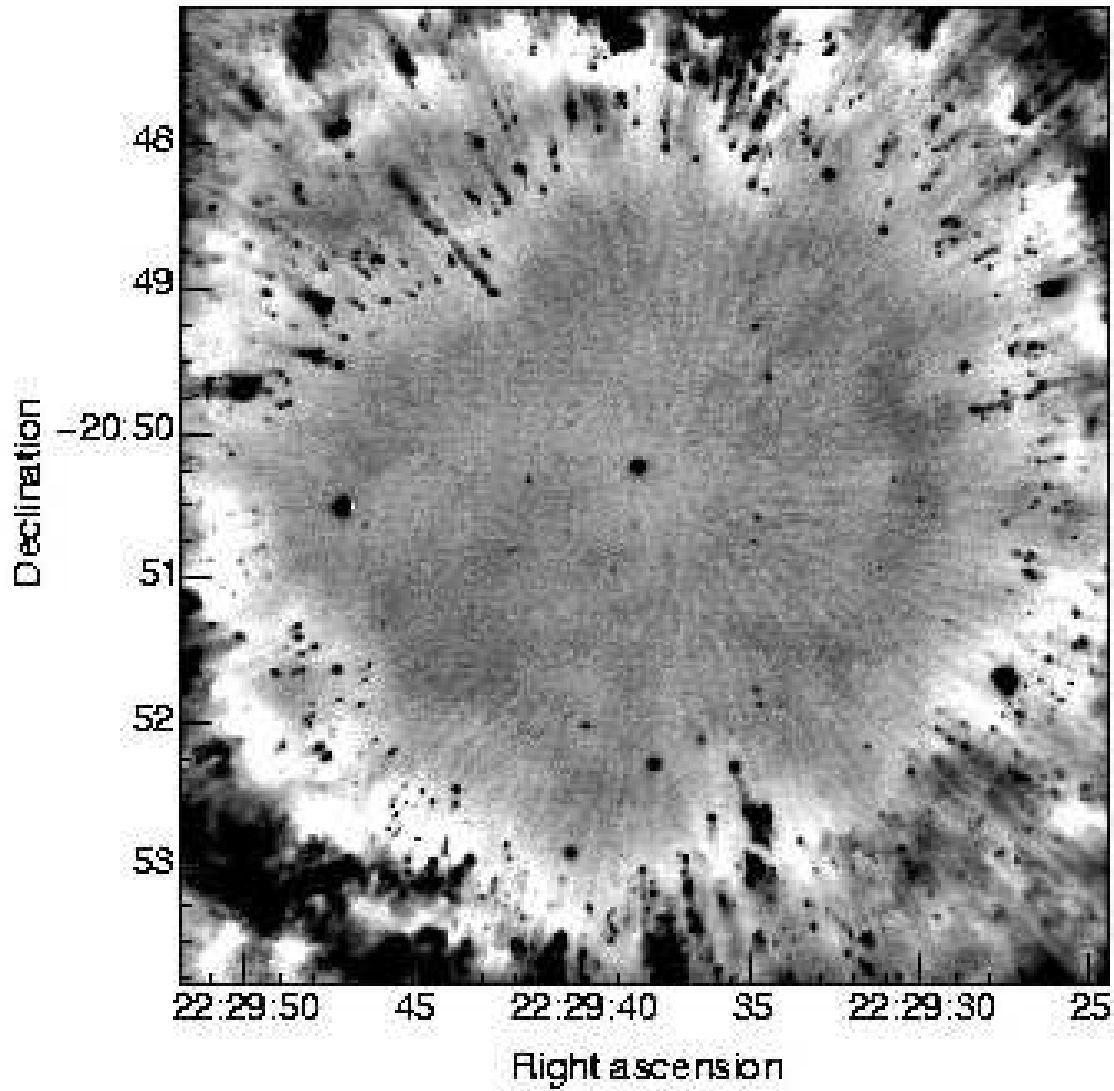}}
\caption{A high contrast, negative greyscale represenation
of an \ha\ + \nii\ image of the core of NGC 7293. The radial
`spokes' that are prominent in the outer regions of the bright
nebulosity can be seen to continue faintly to within 30\arcsec\ of
the central star (epoch 2000).}
\label{reffig10}
\end{figure*}

\begin{figure*}
\epsfclipon
\centering
\mbox{\epsfysize=4in\epsfbox[20 20 575 371]{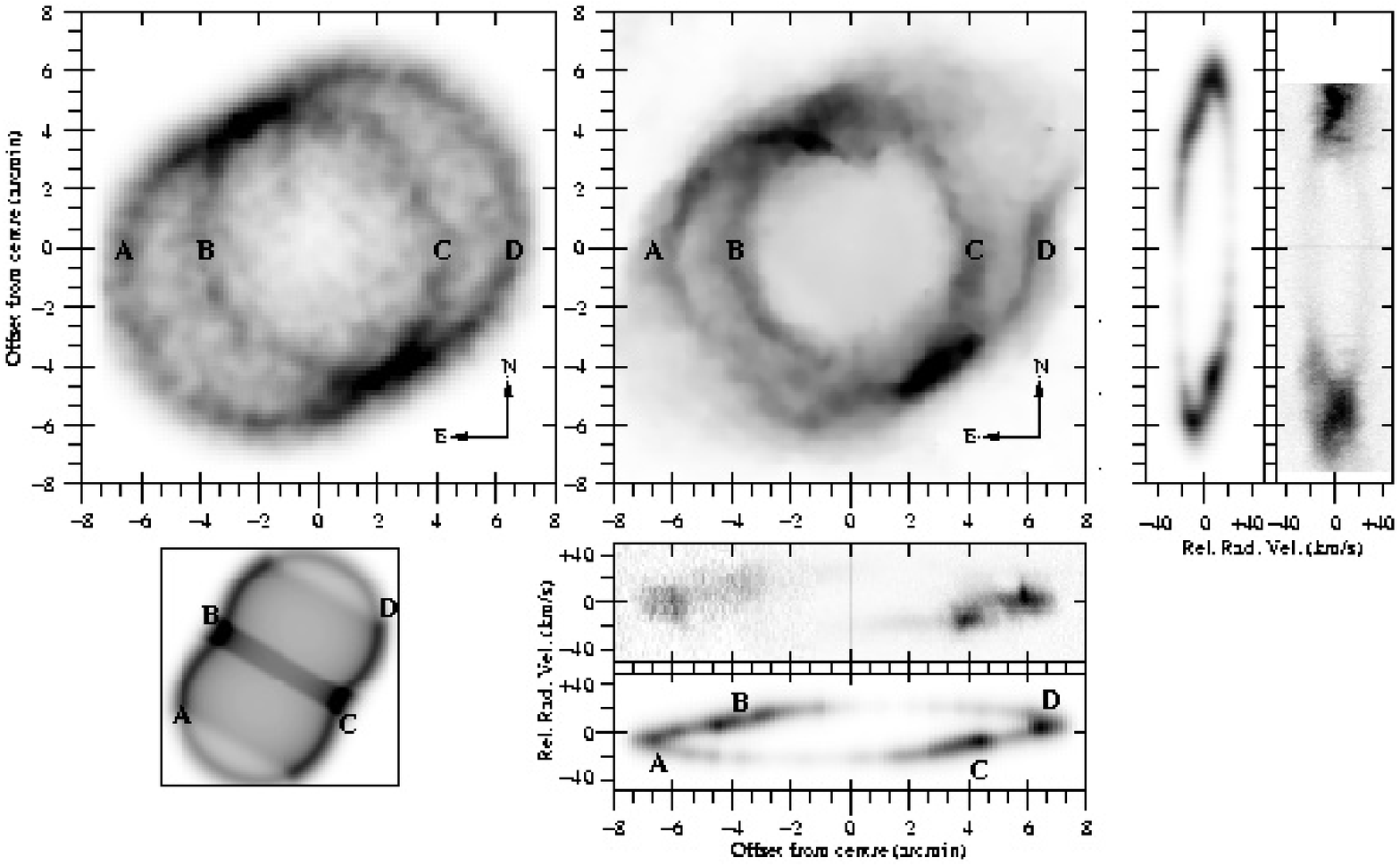}}
\caption{The image of the bright helical \niil\ emitting rings 
and the observed E--W and N--S pv arrays of \niil\ profiles are
simulated using the XSHAPE code. An EW section through 
the bi-polar model that was used is shown (bottom left box) with a central
toroidal ring (viewed here from below). 
The synthetic image (top left box) should be compared with
the observed image (top middle box).
Likewise the synthetic pv array to the left of the observed
NS one (top right boxes) and the synthetic array below the observed
one (bottom right boxes) should be compared. The velocities
for the key positions A--D in Fig. 6 match convincingly.}
\label{reffig10}
\end{figure*}

\section*{Acknowledgements}

The authors wish to thank the staff of the AAT (Australia),
SPM telescope (Mexico), NTT (La Silla, Chile), and Skinakas
telescope (Greece)  for their help
during these  observations.
Skinakas Observatory is a collaborative project of the University of
Crete, the Foundation for Research and Technology-Hellas and
the Max-Planck-Institut f\"ur Extraterrestrische Physik.JAL is grateful
to the Royal Society for financing his July 2004 stay in Manchester
and gratefully acknowledges financial support from CONACYT (M\'ex)
grants 32214-E and 37214 and DGAPA-UNAM IN114199. We would like to thank
Peter Van Hoof for helpful discussions on the \heiil\ fine structure
relative strength calculations.

\label{lastpage}
\end{document}